\documentclass[12pt,preprint]{aastex}

\usepackage{natbib}
\usepackage{xspace}
\usepackage{amssymb}
\usepackage{amsmath}
\usepackage{tabularx}
\usepackage{graphicx}
\newcommand{\um}{\ensuremath{\mu\mbox{m}}\xspace}
\def\cdrev#1{#1}
\newcommand{\km}{\,\hbox{km}\xspace}
\newcommand{\AU}{\,\hbox{AU}\xspace}
\newcommand{\arad}{\ensuremath{a_{\rm rad}}\xspace}
\newcommand{\coll}{{\rm coll}}
\newcommand{\PR}{{\rm PR}}
\newcommand{\taucoll}{\ensuremath{\tau_{\rm coll}}\xspace}
\newcommand{\taucomet}{\ensuremath{\tau_{\rm comet}}\xspace}
\newcommand{\rgain}{\ensuremath{{\cal R}_{\rm gain}}\xspace}
\newcommand{\rloss}{\ensuremath{{\cal R}_{\rm loss}}\xspace}
\newcommand{\nvis}{\ensuremath{n_{\rm vis}}\xspace}

\newcommand{\rhat}{\ensuremath{\hat r}\xspace}
\newcommand{\vcoll}{\ensuremath{v_{\rm coll}\xspace}}
\newcommand{\sigcoll}{\ensuremath{\sigma_{\rm coll}}\xspace}
\newcommand{\sigtot}{\ensuremath{\sigma_{\rm tot}}\xspace}
\newcommand{\tsweep}{\ensuremath{t_{\rm s}}\xspace}
\newcommand{\vkep}{\ensuremath{v_{\rm k}}\xspace}
\newcommand{\Ncomet}{\ensuremath{N_{\rm c}}\xspace}
\newcommand{\mcomet}{\ensuremath{m_{\rm c}}\xspace}
\newcommand{\ncoll}{\ensuremath{n_{\rm coll}}\xspace}
\newcommand{\nPR}{\ensuremath{n_{\rm PR}}\xspace}
\newcommand{\rhocomet}{\ensuremath{\rho_{\rm c}}\xspace}
\newcommand{\acomet}{\ensuremath{a_{\rm c}}\xspace}
\newcommand{\ablowout}{\ensuremath{a_{\rm b}}\xspace}
\newcommand{\fdust}{\ensuremath{f_{\rm d}}\xspace}
\newcommand{\avis}{\ensuremath{a_{\rm vis}}\xspace}
\newcommand{\mvis}{\ensuremath{m_{\rm vis}}\xspace}
\newcommand{\tauPR}{\ensuremath{\tau_{\rm PR}}\xspace}
\newcommand{\tausubl}{\ensuremath{\tau_{\rm subl}}\xspace}
\newcommand{\taukep}{\ensuremath{t_{\rm kep}}\xspace}
\newcommand{\Mstar}{\ensuremath{M_{\star}}\xspace}

\newcommand{\Msun}{\ensuremath{M_{\odot}}\xspace}
\newcommand{\Mearth}{\ensuremath{M_{\oplus}}\xspace}
\newcommand{\Mdisk}{\ensuremath{M_{\rm disk}}\xspace}
\newcommand{\yr}{\mbox{yr}\xspace}
\newcommand{\rmd}{\ensuremath{\mbox d}}

\slugcomment{Submitted to the Astrophysical Journal}

\shorttitle{Age dependence of the Vega Phenomenon: Theory}
\shortauthors{Dominik \& Decin}

\begin{document}


\title{The age dependence of the Vega phenomenon: Theory}


\author{C. Dominik}
\affil{Sterrenkundig Instituut `Anton Pannekoek', Kruislaan 403,
  NL-1098 SJ Amsterdam, The Netherlands}
\email{dominik@science.uva.nl}

\and

\author{G. Decin}
\affil{Instituut voor Sterrenkunde, KU Leuven,
Celestijnenlaan 200 B, B-3001 Leuven, Belgium}
\email{greet@ster.kuleuven.ac.be}

\begin{abstract}
  In a separate paper \citep[][henceforth paper I]{decin-I}, we have
  re-examined the observations of IR excess obtained with the ISO
  satellite and discussed the ages of stars with excess.  The amount
  of dust (measured by the luminosity fraction
  $\fdust=L_\mathrm{IR}/L_{\star}$) seen around main-sequence stars of
  different ages shows several interesting trends.  To discuss these
  results in the context of a physical model, we develop in this paper
  an analytical model for the dust production in Vega-type systems.
  Previously it has been claimed that a powerlaw slope of about -2 in
  the diagram plotting amount of dust versus time could be explained
  by a simple collisional cascade.  We show that such a cascade in
  fact results in a powerlaw $\fdust\propto t^{-1}$ if the dust
  removal processes are dominated by collisions.  A powerlaw
  $\fdust\propto t^{-2}$ only results when the dust removal processes
  become dominated by Pointing-Robertson drag.  This may be the case
  in the Kuiper Belt of our own solar system, but it is certainly not
  the case in any of the observed disks.  A steeper slope can,
  however, be created by including continuous stirring into the
  models.  We show that the existence of both young and old Vega-like
  systems with large amounts of dust ($\fdust\simeq 10^{-3}$) can be
  explained qualitatively by Kuiper-Belt-like structures with
  \emph{delayed stirring}.  Finally, the absence of young stars with
  intermediate amounts of dust may be due to the fact that stirring
  due to planet formation may not be active in young low-mass disks.
  The considerations in this paper support the picture of simultaneous
  stirring and dust production proposed by
  \citet{2002ApJ...577L..35K}.
\end{abstract}



\section{Introduction}
\label{sec:introduction}

\cdrev{Debris disks are thought to form after the depletion of gas in young
circumstellar disks.  During the first $\sim10$~Myr of a star's life,
small dust grains grow by coagulation and finally produce
planetesimals \citep{1993ARA&A..31..129L}.  When the average velocity
of the collisions increases due to stirring by planets or large
planetesimals forming in the disk, the encounters between the
planetesimals become destructive, initiating a collisional cascade.
Large quantities of small dust grains are produced in such a
cascade.  These grains emit infrared and submm radiation, making the
debris disk visible \citep[e.g.][]{Aum-ea-84,zucker}.}

Since most Vega-like excess stars are spatially unresolved by infrared
telescopes, the quantity used to measure the amount of dust is the
\textit{fractional luminosity}, \fdust: the ratio of the dust emission
$L_\mathrm{IR}$ and the stellar luminosity $L_{\star}$
\begin{equation}
\label{eq:35}
\fdust = \frac{L_\mathrm{IR}}{L_{\star}} \quad.
\end{equation}
\fdust can be interpreted as a \textit{covering fraction}.
It measures the fraction of the sky seen from the star which is
covered by dust, and therefore the fraction of the stellar radiation
which will be absorbed and reprocessed to the infrared.

Studying the age dependence of the Vega phenomenon has been at the
focus of several studies using the ISO satellite.  \cdrev{The Vega phenomenon
appears to be much more widespread around younger stars than around
older stars.  Studying a volume-limited sample of stars near the Sun,
\citet{1999Natur.401..456H,2001A&A...365..545H} found that more than
60\% of the observed stars below an age of 400 Myrs show a Vega-like
excess, while only 9\% of older stars can be classified as Vega-like.
This result is also confirmed by submm observations
\citep[e.g.][]{zucker,2003MNRAS.342..876W}.  Looking at a few selected
vega-like stars, \citet{2000EM&P...81...27K} noted a $\sim t^{-1}$
decline.  Studying members of young clusters with different ages,
\citet{2001ApJ...555..932S} found a trend in the average amount of
dust seen around stars of different ages which was described by a
powerlaw dependence of \fdust versus time.  Spangler et al.
tentatively interpreted this result as a global trend due to a
collisional cascade.}

A thorough study of the required parameters, age and fractional
luminosity, was done in paper~I to re-examine the available ISO data.
It led to the following implications for models: {\sl (i)} debris
disks are more common around young stars than around old ones, but
{\sl (ii)} a general power-law for the dust mass versus age with slope
of $\sim -2$, as found by e.g. \citet{holland98:_submil} and
\citet{2001ApJ...555..932S} could not be confirmed.  \cdrev{Two main
  issues are causing the difference in results.  There clearly is a
  strong decrease in circumstellar material from weak-line T Tauri
  stars to the youngest main sequence stars, but we reject weak-line
  T~Tauri stars as debris disks \citep{lagrange00:PPIV} because these
  disks are likely to still be gas rich and governed by different
  physics \citep{1996rdfs.conf..137Asa}.  Also, we find a significant
  number of old stars with large \fdust-values, contradicting a global
  powerlaw decrease of \fdust.}  There may still be a decline of dust
mass for the youngest stars, with slope of $\sim$ -1.3, but surely not
for the older ones.  {\sl (iii)} We found that the maximum excess does
not depend on age, also demonstrating the absence of a global
declining trend applying to all stars.  A final interesting result
from paper~I is that {\sl (iv)} there is an apparent scarcity of young
stars with intermediate or low infrared excesses.

Little modeling on the general time evolution of Vega-like disks has
been done so far.  The collisional evolution of the solar system
Asteroid Belt has been modeled extensively, mainly to understand the
observed present day size and rotation distribution of large
asteroids.  Similar models for Kuiper Belt objects have been described
by \citet[e.g.][]{1997Icar..125...50D}.  Production of dust in the
Asteroid Belt has been studied by \citet{1997Icar..130..140D},
focusing on the detailed size distribution produced by the collisional
cascade at small sizes, which is influenced by both collisions and
interaction with radiation.  Studies of collisional cascades in the
Kuiper Belt have also considered the formation of dust, but only in
very basic ways \citep{1997ApJ...490..879S}.  The most detailed
numerical models of combined growth and destruction of bodies in disks
around stars are recent calculations by
\citet{kenyon99:_accret_early_kuiper_belt,kenyon99:_accret_early_outer_solar_system}
and
\citet{2001AJ....121..538K,2002AJ....123.1757K,2002ApJ...577L..35K}
who study the evolution of the outer protoplanetary disk.  We will get
back to these papers in the discussion.

The goal of the present paper is to better understand the age
dependence of the dust mass in Vega-type stars in terms of a simple
model.  Therefore, an analytical model is developed in Sect.~2 which
describes the evolution of debris dust.  This model is then confronted
with the observations and discussed in Sect.~3.

\section{Collisional model}
\label{sec:collisional-model}

We will use a very simple collisional model to derive the amount of
dust produced by a collisional cascade of large bodies, and to compute
the decay of the dust present in the system with time.  A similar
model has been used to estimate the lifetime of Vega-like disks by
\citet{2001AJ....121..538K}, but we go further and compute the amount
of dust as a function of time under the assumption that the evolution
of the disk is purely due to a collisional cascade.  In deriving the model,
we will make use as much as possible of general laws governing all
kinds of collisions, and refer as little as possible to detailed
material properties.  The underlying model will be that of the Kuiper-
or Asteroid belt of our own system:  a number of large bodies,
colliding to produce and replenish the dust we see.

\subsection{Collisional removal of comets}
\label{sec:coll-remov-comets}
Let us assume that all the dust in the system is produced as the end result
of a collisional cascade involving large bodies.  We therefore start
with a number $\Ncomet$ of comets.  These comets have a radius
\acomet, a geometrical cross-section for collisions between comets of
$\sigcoll=4\pi\acomet^2$, and a mass $\mcomet=\frac{4\pi}{3}\rhocomet
\acomet^3$, where \rhocomet is the density of the cometary material.

For the collisions between the comets, we use a particle-in-a-box
model:  all comets are moving through a given volume $V$ which could
be an entire planetary disk or just a limited region like the Asteroid
or Kuiper Belt.   We assume that the occupied volume is a section
of a wedge-shaped disk between distances $r_1$ and $r_2$ with a local
half-height $H(r)=h \cdot r$ where $h$ is the normalized height $H/r$ of
the disk.  The volume of such a wedge segment is
\begin{equation}
\label{eq:2}
V=\frac{4\pi}{3} h \left( r_2^3 - r_1^3 \right) 
=: \frac{4\pi}{3}h \rhat^3 \quad,
\end{equation}
where \rhat is a number close to $r_2$ if r$_1$ is not too close to
$r_2$.  If we use numbers typical for the Kuiper Belt ($r_1\simeq
35\AU, r_2\simeq 50 \AU$,
$h\simeq .2$) we find $\rhat=43.5\AU$ and $V=6.25\times 10^{44}$cm$^3$.

The particle-in-a-box model assumes that there is a fixed collision
velocity between different comets, which we denote with \vcoll.  For
simplicity, we will write this collision velocity as a fraction $\nu$
of the Kepler velocity:
\begin{equation}
\label{eq:3}
\vcoll = \nu \vkep(\rhat) = \nu \,\sqrt{\frac{GM}{\rhat}} \quad,
\end{equation}
where $G$ is the gravitational constant and $M$ is the mass of the star.
One can easily see that $\nu$ is related to the relative height $h$ of
the disk.  Collisions between bodies in a disk are due to differences
in the inclinations and eccentricities of orbits.  A particle on a circular
orbit with inclination $i$ will cross the midplane of the system with
a vertical speed $\vkep\times\tan i$.  Therefore,  a disk with a
normalized height of $h=\tan i$ will show typical collision velocities
$h\times \vkep$.  We can therefore assume $\nu=h$ whenever we
numerically evaluate expressions.

With the collision velocity $\vcoll$ and the volume given, we can
compute the \emph{sweeping time} \tsweep{} which is the time needed by
a single comet to sweep the entire volume accessible to the cometary
cloud
\begin{equation}
\label{eq:4}
\tsweep = \frac{V}{\vcoll \sigcoll} \quad.
\end{equation}
Note that the sweeping time is actually independent of the collision
velocity, since the volume occupied by the cometary cloud also grows
with $\vcoll$.  The collision time for comets is then given by
\begin{equation}
  \label{eq:33}
  \taucomet = \frac{\tsweep}{\Ncomet} \quad.
\end{equation}

For dust production to occur, the collisions between comets must be
destructive.  Assuming that any such collision removes 2 comets from
the cloud, we get for the time evolution of the number of comets
\begin{equation}
\label{eq:5}
\frac{d\Ncomet}{dt} = -2 \frac{\Ncomet^2}{\tsweep} \quad,
\end{equation}
with the solution
\begin{equation}
\label{eq:6}
\Ncomet(t) =\frac{\Ncomet(0)}{1+2\Ncomet(0)t/\tsweep} \quad.
\end{equation}

For small times, the number of comets will be constant while for long
times we find
\begin{equation}
\label{eq:7}
\Ncomet(t) \simeq \frac{\tsweep}{2t} 
\quad\mbox{for}\quad
t\gg \frac{\tsweep}{2\Ncomet(0)} \quad.
\end{equation}

The number of comets after a given time is therefore initially
constant and later turns into a powerlaw with slope $-1$.

\subsection{Dust production}
\label{sec:dust-production}
Little is known about the dust production in collisions between comets
or asteroids, as the dust production rate will depend to some extent
on the internal structure of the body.  If the body has already been
fractured many times in previous collisions and now basically is a
\emph{bag of sand}, a massive collision will release large amounts of
dust.  If, on the other hand, the body still has some internal
strength, it can be expected to produce a distribution of fragment
sizes.  While the experimentally determined distributions probable
give good results for the larger fragments, the production rate of
dust grains is very uncertain.  Therefore we follow a different road
and assume that dust production is the result of a collisional cascade
where increasingly small particles collide with each other to produce
smaller and smaller fragments.  It has been known for many years that
the size distribution produced by such a process assumes the shape of
a powerlaw with the slope $f(m)\propto m^{-q}$ or (for spherical
particles) $f(a)\propto a^{2-3q}$ where $q$ can be shown to be equal
to $11/6$ for a self-similar collisional cascade, largely independent
of material properties
\citep{1968IAUS...33..486D,1994Icar..107..117W,1996Icar..123..450T}.
Significant deviations from this law appear only when the gravity
component of the impact strength is considered - but for bodies
smaller than 1km, the binding energy is dominating.  The steady-state
powerlaw is based on the competition between production and removal
processes.  At the smallest sizes, this equilibrium will be disturbed
by the non-collisional grain removal processes like radiation pressure
(blowout) and Poynting-Robertson drag \citep{1997Icar..130..140D}.
Also collisions with $\beta$-meteorites may play a role
\citep{1996rdfs.conf..137Asa,2000ApJ...539..424K}.  The disturbance in
the powerlaw can actually progress like a wave to larger particle
sizes \citep{1997Icar..130..140D}.  For the simple estimates 
in this paper, we are going to ignore these additional
processes.  We will assume that a powerlaw distribution is produced by
the collisions, and that the distribution continues with the same
slope down to the particles which are removed from the system by
another process.  We write the size distribution as
\begin{equation}
\label{eq:13}
f(a) = f_a a^{\gamma}
\end{equation}
where $\gamma=2-3q$ is equal to $-3.5$.  
It is easy to show that the collisional timescales are slowest at the
large-particle end of the distribution.  Then, the collisions between
large comets determines the injection of material in the cascade and
therefore directly the dust production rate at the small particle end
of the cascade.  The steady state dust production rate $\rgain$ in
such a model is proportional to the number of comet collisions per
time, i.e.
\begin{equation}
\label{eq:16}
\rgain = k_1 \Ncomet^2 \quad,
\end{equation}
where $k_1$ is a constant.

Since the dust production rate is directly fixed by the collisional
cascade, the dust removal processes govern the amount of dust visible
in a given source.

\subsection{Dust removal}
\label{sec:dust-removal}
Before we look into the details of dust removal, we can already
provide the general time dependence of the dust content in a debris
disk driven by a purely collisional evolution.  From Eq.~\eqref{eq:16}
it is clear that the dust production rate is proportional to rate of
collisions between comets, thus $\propto \Ncomet^2$.  Since the number
of comets decreases as $t^{-1}$, the dust production rate is
proportional to $t^{-2}$. However, we cannot conclude from this that
the amount of dust present in steady state shows the same time
dependence - the mechanism which is responsible for grain losses has
to be considered as well.  If grain losses are due to collisions among
the visible grains themselves, the loss rates will generally be
proportional to the number of visible grains squared.  However, if
some other process removes the visible grains, the loss rates will
only linearly depend on the number of the grains.  Thus we have
\begin{equation}
\label{eq:17}
\rloss = k_2 \nvis^{\alpha} \quad,
\end{equation}
where $\alpha$ is either 1 or 2.  In steady state we have
\begin{equation}
\label{eq:18}
\rgain = \rloss \quad,
\end{equation}
and therefore
\begin{equation}
\label{eq:19}
\nvis = \left( \frac{k_1}{k_2} \right)^{1/\alpha} \Ncomet^{2/\alpha}
\quad.
\end{equation}

Thus, for internal collisional processes dominating the grain loss, we
can expect the number of visible grains to be proportional to the
number of comets left in the system, corresponding to a $t^{-1}$
dependence of the dust amount.  If other processes dominate grain
removal, a $\Ncomet^2$ (corresponding to $t^{-2}$) dependence should
be expected.

In the following we will show under what circumstances which
process dominates, and therefore which powerlaw we should expect from
the observations.

\subsubsection{Poynting-Robertson lifetime}
\label{sec:poynt-roberts-lifet}

\cdrev{Poynting-Robertson drag removes angular momentum from particles in
orbit around a star and causes them to spiral inward.}  The
Poynting-Robertson drag reduces the size of a circular orbit according
to \citep{burns79:_radiat}
\begin{equation}
\label{eq:8}
{\dot r} = - \frac{G \Mstar}{r} \frac{2\beta}{c} 
= -2 \vkep^2 \frac{\beta}{c} \quad,
\end{equation}
where $c$ is the speed of light and $\beta$ is the ratio of radiative
and gravitational acceleration of a particle.  \cdrev{The radiative
  force acting on the particle depends on the absorption and scattering
  cross sections \citep{burns79:_radiat}.  For the simple estimates in
  our study, we do not wish to discuss the detailed grain properties.
  Instead, we limit ourselves to large grains (compared to the
  wavelength of the radiation emitted by the star) and assume the
  grains to be perfect absorbers.  In this limit, the absorption cross
  section of the particle is $\pi a^2$ and scattering does not
  contribute to the radiation pressure \citep{1981lssp.book.....V}.
  $\beta$ is then inversely proportional to the particle size, so we
  can write \citep{backman-review}
\begin{equation}
\label{eq:9}
\beta(a) = 
\frac{3L_{\star}}{16\pi cG\Mstar a \rhocomet} =
\frac{1}{2}\frac{\ablowout}{a}
\end{equation}
where} \ablowout{} is the blowout size, i.e. the size of particles which
will be blown out of the system by radiation pressure.  We define
these to be the particles with $\beta=1/2$.  Particles with
$\beta=1/2$ will leave the system when ejected in a collision between
particles unaffected by radiation pressure \citep{burns79:_radiat}.

The PR timescale of particles of radius $a$ moving in orbits with a
semi-major-axis $r$ is then given by
\begin{align}
\label{eq:10}
\tauPR(a) &= \frac{r}{\dot r} = \frac{r^2c}{2 G \Mstar} \frac{a}{\ablowout}\\
&= 2\cdot 10^6 \yr \left( \frac{r}{50\AU} \right)^2
\frac{\Msun}{\Mstar} \frac{a}{\ablowout}
\quad.
\end{align}

\cdrev{Poynting-Robertson drag is frequently used in the literature to
  compute the dust loss in debris disks
  \citep[e.g.][]{1998ApJ...505..897J,1999A&A...350..875J,2003ApJ...590..368L}.
  It certainly provides an upper limit on removal time scales.
  However, as we will show below, collisions will always dominate the
  removal process in the currently observable debris disks.}

\subsubsection{Collisional lifetime}
\label{sec:collisional-lifetime}

The collisional cross-section between two particles with radii $a_1$
and $a_2$ is given by
\begin{equation}
\label{eq:11}
\sigma(a_1,a_2) = \pi (a_1+a_2)^2 \quad.
\end{equation}
Of course, not all collisions will be destructive - only collisions
with particle sizes above a certain threshold can destroy the bigger
particle.  We denote by $\varepsilon a_1$ the size of the smallest
particle which can (at a given collision velocity) still destroy a
target particle with radius $a_1$.  The total destructive collision
cross section provided by the particle size distribution in order to
destroy particles with radius $a_1$ is then given by
\begin{align}
\label{eq:12}
\sigtot(a_1) &= \int\limits_{\epsilon a_1}^{\acomet} f(a) \pi
(a_1+a)^2 \rmd a\\
&= \int\limits_{\epsilon a_1}^{\acomet} f_a a^{\gamma} \pi (a_1+a)^2
\rmd a
\quad.
\end{align}
In a powerlaw distribution with $\gamma<-3$, the cross section will be
dominated by collisions with grain sizes $\epsilon a_1$:
\begin{equation}
\label{eq:14}
\sigtot(a) \simeq -f_a \frac{\pi\epsilon^{\gamma+1}}{\gamma+1}
a^{\gamma+3}
=: \epsilon_0 f_a a^{\gamma+3}
\end{equation}
where we have defined
\begin{equation}
\label{eq:37}
\epsilon_0=-\pi\epsilon^{\gamma+1}/(\gamma+1) \quad.
\end{equation}

The collisional lifetime due to collisions with other particles in the
cascade is therefore given by
\begin{equation}
\label{eq:15}
\taucoll(a) = \frac{V}{\vcoll \sigtot(a)}
= \frac{V}{\nu \vkep} \frac{1}{\epsilon_0 f_a a^{\gamma+3}}
\end{equation}
which for $\gamma=-3.5$ is proportional to $\sqrt{a}$.  However, the
absolute value of the timescale depends on the normalization constant
$f_a$ of the equilibrium size distribution.

We still need to compute $\epsilon$.  Collisions are destructive if
the kinetic energy of the collision is approximately equal to the
binding energy of the two bodies.  In general, the binding energy is
composed of a component due to the material properties, and a
gravitational component.  However, for bodies with 1 km or less in
diameter, the gravitational component can be ignored.  For a collision
to be destructive we therefore require
\begin{equation}
  \label{eq:28}
  \frac{1}{2}\mu \vcoll^2 = S (m_1+m_2)
\end{equation}
where $\mu$ is the reduced mass and $S$ is the binding energy per
mass.  We then ask what the minimum mass $m_2$ is which can still
destroy a body with mass $m_1$ in a collision with a given velocity.
Using $m_2 = \epsilon^3 m_1$ we find
\begin{equation}
  \label{eq:29}
  \epsilon^3 = \frac{\vcoll^2}{4S} -
   \sqrt{\frac{\vcoll^4}{16S^2} - \frac{\vcoll^2}{2S}}
  - 1 \quad.
\end{equation}
Note that $\epsilon$ contains the only important
velocity dependence of our estimates.  The higher the collision
velocity, the larger the size interval of destructive impactors, and
the larger the destruction rates of a given particle size.  We will
use a binding energy of $S=2\cdot 10^6$\,erg/s, \cdrev{a value typical for the
icy bodies in the outer solar system
\citep{kenyon99:_accret_early_kuiper_belt}.  Table~\ref{tab:eps0}
lists the value of $\epsilon_0$ as a function of $S$ and $\vcoll$.}

\placetable{tab:eps0}

\subsubsection{Sublimation}

\cdrev{Sublimation of ice can be a removal process for small grains in the
inner regions of the disk.  Ice sublimation is an extremely strong
function of the grain temperature.  In a range between 90 and 120\,K,
the sublimation time $\tausubl$ can be approximated by the following
expression \citep{backman-review}
\begin{equation}
\label{eq:36}
\tausubl \approx 10^6 \frac{a}{1\um} \left( \frac{T_g}{100\,K}
\right)^{-55} \mbox{yr}
\end{equation}
where $T_g$ is the grain temperature.  Vega-like disks show emission
which is usually dominated by 60\um excess, only very few stars show
detectable access at 25\um \citep[e.g.][]{2002A&A...387..285L}.  The
disk emission usually peaks around 60\um or at even longer wavelengths
\citep[e.g.][]{zucker}.  Using Wiens law, this corresponds to dust
temperatures below 100\,K.  The sublimation times for such
temperatures are generally comparable with or longer than
Poynting-Robertson drag timescales, even for the A stars among
Vega-like stars (e.g. $\beta$ Pic, $\alpha$ Lyra and $\alpha$ Psa, see
\citet{backman-review}).  In stars of later spectral type, grains at
similar distance from the star will be colder.  While grain
sublimation may play a role in setting the inner boundaries of debris
disks, it does not dominate the removal processes. We therefore ignore
grain sublimation in the following.}

\subsection{Steady-state size distribution}
\label{sec:steady-state-size}

In steady state in a collisional cascade, $f_a$ can be determined
equating the mass flux through the collisional cascade with influx of
new material at the top end.  The mass flux through the chain will in
general depend on the details of the collision physics, but for the
purpose of this study, a simple estimate is sufficient.  The
collisional mass loss of particles with sizes in a scale-free size
interval between $a$ and $2a$ is given by
\begin{align}
\label{eq:20}
\dot{M}(a) = \frac{f(a) a \frac{4\pi}{3}\rhocomet a^3}{\taucoll(a)}
= f_a^2 a^{2\gamma+7} \frac{\frac{4\pi}{3}\vcoll\rhocomet\epsilon_0}{V}
\quad.
\end{align}
For the equilibrium slope $\gamma=-3.5$, this value is independent of
grain size.  This mass loss is an excellent estimate for the mass flux
through the cascade if in a typical collision the largest fragment has
about half the diameter of the impactors.  Equating $\dot{M}(a)$ with
the inflow of mass into the collisional cascade (Eq.~\eqref{eq:5}
multiplied by the \mcomet), we find
\begin{equation}
\label{eq:21}
f_a = \Ncomet \sqrt{\frac{2\mcomet\sigcoll}{\frac{4\pi}{3}\rhocomet
    \epsilon_{0}}} = 
\Ncomet \acomet^{5/2} \sqrt{\frac{8\pi}{\epsilon_{0}}}
\quad.
\end{equation}

\subsection{Dust visible in steady state}
\label{sec:dust-visible-steady}

The dust visible in the steady-state solution of a collisional cascade
consists of two components.  One component are small dust grains which
are being blown away by stellar radiation (the \emph{wind grains}).
Each of these grains contributes to the radiation (scattering and
IR/submm emission) only for the duration of about 1 Kepler time,
typically a few hundred years.  The second component consists of
grains which are too large to be blown away, but are already small
enough to provide significant surface for interaction with photons
(the \emph{orbiting grains}).  Such grains contribute to the radiation
as long as they live in the given orbit, i.e. for a collision time or
a PR-Drag time, whatever is shorter.  The relative importance of both
components can be easily estimated.   The critical size for
blowout of grains is $\ablowout$.  Of the grains smaller than this
size, there will be a size $\arad$ which will dominate the radiation
interaction of the grains below the blowout limit.  We assume that the
destruction of bigger particles produces the two different sizes
according to the steady state powerlaw, i.e. the small grains are
produced more frequently by factor $(\ablowout/\arad)^{-\gamma}$.  The
\emph{production of grain surface} (which is most important for the
visibility of dust) proceeds then with a ratio
$(\ablowout/\arad)^{-\gamma-2}$.  Therefore, the orbiting grains will
dominate if their lifetime $\tau$ meets the condition
\begin{equation}
\label{eq:23}
\tau \ge (\ablowout/\arad)^{-\gamma-2} \taukep \quad.
\end{equation}
For a debris disk around a main-sequence star, the blowout size is
typically $1-10\um$ (depending upon spectral type
\citep{1988ApJ...335L..79A}) while $\arad$ 
should be of the order of $1\um$.  Therefore, as long as the life time
of the smallest orbiting grains is at least 10--100 orbits, orbiting
grains will always dominate the visibility of debris disks.  In the
following discussion, we will therefore focus on the steady-state
abundances of grains with sizes just above the blowout size.  We will
call this size of the visible dust grains $\avis$.

\subsection{Steady-state abundance of grains}
\label{sec:steady-state-abund}

The number of grains visible in the disk at a given time will be
determined by the dominating grain-loss mechanism.

\subsubsection{Collisionally-dominated grains}
\label{sec:coll-domin-grains}

If the visible dust grains are still being removed by
collisions and have their steady-state abundance, we find
\begin{equation}
\label{eq:24}
\ncoll = f(\avis) da = f_a a^{\gamma} da \approx \Ncomet \acomet^{5/2}
\sqrt{\frac{8\pi}{\epsilon_0}}\avis^{\gamma+1}
\end{equation}
where we have again assumed $da\approx a$.

\subsubsection{PR drag dominated grains}
\label{sec:pr-drag-dominated}

If the visible grains are removed by PR drag, this means that the
collisional cascade is effectively terminated at the size $\avis$.
Then, the number of grains can be estimated by equating the mass flux
through the collisional cascade with PR driven losses.  Therefore we
have
\begin{equation}
\label{eq:22}
\nPR = \frac{\mcomet}{\mvis}  \dot{\Ncomet} \tauPR(\avis) = \Ncomet^2
\frac{3\acomet^5 c}{\avis^3 r^2\vkep} \quad.
\end{equation}

\subsubsection{General case}
The number of grains visible in the disk will then be given by
\begin{equation}
\label{eq:25}
\nvis = \min(\ncoll,\nPR) \quad.
\end{equation}
and the covering fraction $\fdust$ of the disk is
\begin{equation}
  \label{eq:27}
  \fdust = \nvis Q_{\rm abs}(\avis) \pi \avis^2 \quad,
\end{equation}
where $Q_\mathrm{abs}$ is the usual absorption efficiency of the dust
particles.
At high disk masses, the collisional destruction always dominates the
life time of grains, while at low disk masses, PR drag can become
important.  The disk mass where the switch in processes takes place is
given by the equation $\ncoll=\nPR$. We find
\begin{equation}
\label{eq:1}
M_{\coll\to\PR} = 
\frac{4\pi}{9} \sqrt{\frac{8\pi}{\epsilon_0}}
\frac{\sqrt{\acomet\avis G M_{\star} r^3}\rhocomet{}}{c}
\quad.
\end{equation}

\subsection{Dependencies and typical numbers}
\label{sec:depend-typic-numb}

The equations derived above show many properties of a collisional
cascade and the amount of radiation reprocessing by dust which can be
expected from such a cascade. 
If we insert typical numbers into the above equations, we find the
following results:

\begin{align}
\label{eq:26}
\taucomet &= 7.4\cdot 10^7 \yr
\left( \frac{\rhat}{50\AU} \right)^{3.5}
\frac{\acomet}{1\,\km} 
\frac{\rhocomet}{1 {\rm g\,cm}^{-3}}
\left( \frac{\Msun}{\Mstar} \right)^{0.5}
\frac{10\Mearth}{\Mdisk}
\\
\label{eq:260}
&= 4.9\cdot 10^7 \yr
\left( \frac{\rhat}{50\AU} \right)^{1.5}
\left(
\frac{1\um}{\avis}
\frac{\acomet}{1\,\km} 
\frac{226}{\epsilon_0}
\frac{\Msun}{\Mstar}
\right)^{0.5}
\frac{Q}{1}
\frac{10^{-3}}{\fdust}
\\
\label{eq:30}
\taucoll &= 3.9\cdot 10^2 \yr
\left( \frac{\rhat}{50\AU} \right)^{3.5}
\left( \frac{\acomet}{1\,\km} 
       \frac{1M_{\odot}}{M_{\star}} 
       \frac{\avis}{1\um}
       \frac{226}{\epsilon_{0}}
 \right)^{0.5}
\frac{10\Mearth}{\Mdisk}
\\
\label{eq:31}
\tauPR &= 2.0\cdot 10^6 \yr 
\left( \frac{\rhat}{50\AU} \right)^2
\frac{M_{\odot}}{M_{\star}}
\frac{a}{\ablowout}
\\
\label{eq:32}
M_{\coll\to\PR} &= 1.7\cdot 10^{-3} \Mearth
\left( \frac{\rhat}{50\AU} \right)^{1.5}
\left[
 \frac{226}{\epsilon_0}
 \frac{\acomet}{1\,\km}
 \frac{\avis}{1\um}
 \frac{\Mstar}{\Msun}
\right]^{0.5}
\left( \frac{\rhocomet}{1 {\rm g\,cm}^{-3}} \right)
\\
\label{eq:34}
f_\mathrm{d,coll} &= 2.3\cdot 10^{-3}
\left( \frac{50\AU}{\rhat} \right)^2
\left(
\frac{1\um}{\avis}
\frac{1\,\km}{\acomet}
\frac{226}{\epsilon_0}
\right)^{0.5}
\frac{\Mdisk}{10\Mearth}
\frac{1 {\rm g\,cm}^{-3}}{\rhocomet}
\frac{Q_\mathrm{abs}}{1}
\end{align}
\cdrev{where we have used a value of 226 for the $\epsilon_0$ parameter which
is appropriate for our standard model ($S=2\cdot 10^{6}\,\mbox{erg/g}$,
$\vcoll=0.1\vkep\simeq 0.45\,\mbox{km/s}$).}

First of all, we see that the dust removal timescales through
collisions are much shorter than the removal timescales through PR
drag.  This means that typical debris disks will be collisionally
dominated down to the smallest dust sizes which will then be blown out
by radiation pressure.  Eq.~\eqref{eq:32} shows that the disk mass
(i.e. the mass of all comets in the collisional system) must be as low
as $1.7\cdot 10^{-3}\Mearth$ for the transition to occur.  From
Eq.~\eqref{eq:34} we can see that the \fdust-value at that stage will
typically be $4 \cdot 10^{-7}$ - well below the detection limits of
ISO and IRAS.

\begin{figure}
\plotone{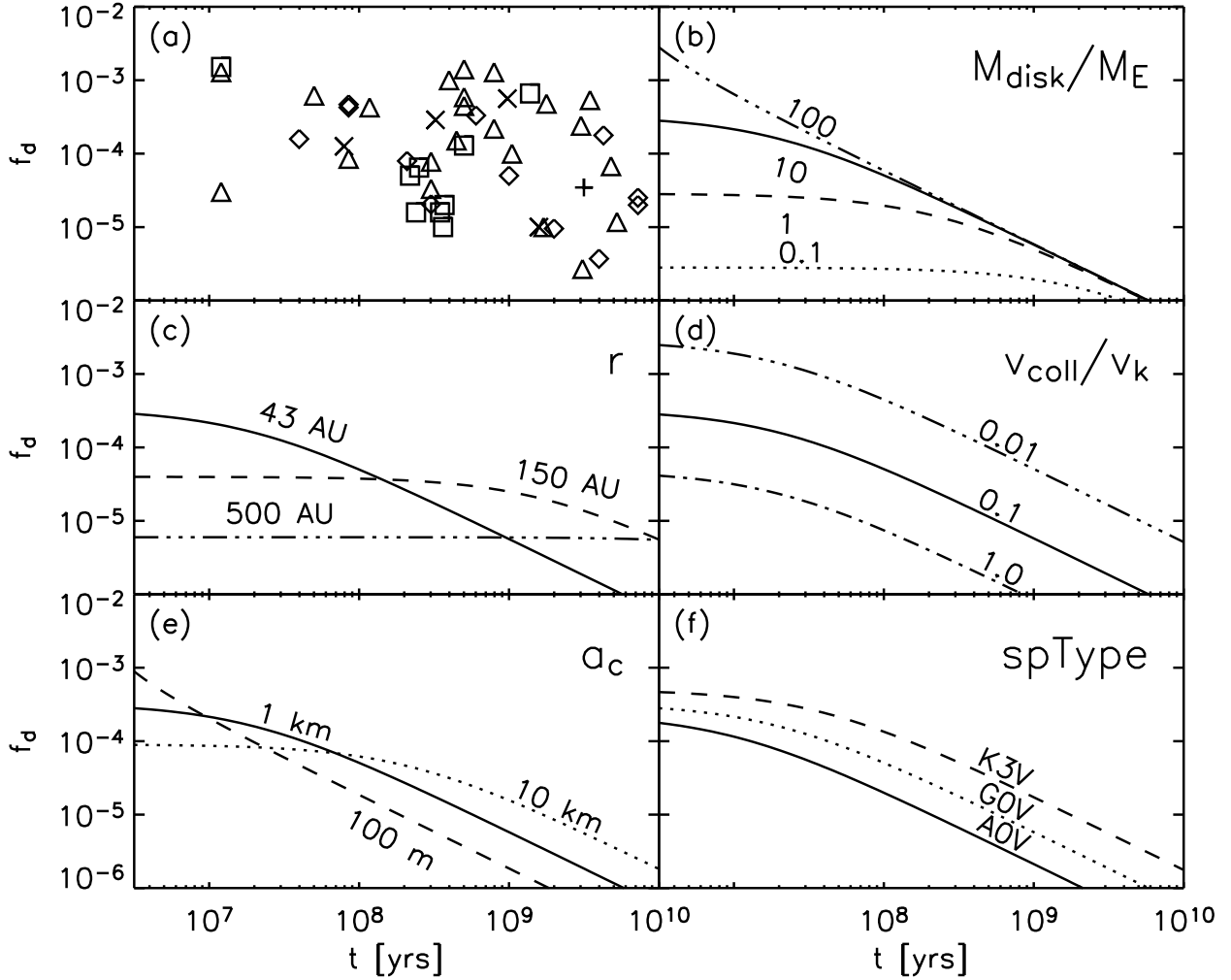}
\caption{Dependence of the fractional luminosity of dust produced by a
  cloud of comets as a function of time.  Panel (a): The observations
  (paper~I).  Explanation of the symbols: $\Box$: A main-sequence
  stars, $\bigtriangleup$: F main-sequence stars, $\Diamond$: G
  dwarfs, $\times$: K dwarfs and $+$: K giant. Panels (b)--(f) show
  the dependence on different parameters, starting from a standard
  model (solid line in all panels).  See text.
  \label{fig:dependence}}
\end{figure}

\placefigure{fig:dependence}

Figure~\ref{fig:dependence} shows the dependence of the dust visible
in the system on the different parameters of the model.  Panel (a)
shows the data taken from paper I.  The solid line in the other five
panels of the figure represents the time dependence of a standard
model with $\Mdisk=10\Mearth$, $\hat{r}=43\AU$, $\nu=0.1$,
$\acomet=1$\,km around an A0 star.  The other lines indicate the
changes due to the variation of a single parameter.  From panel (b) in
Fig.~\ref{fig:dependence}, it can be seen that only the more massive
disks ($>10\Mearth$) reach collisional steady state within
$\taucomet\simeq 10^8$ years.  This is the time when most comets have
seen at least one collision, and when the slope in the \fdust-time
relation turns from zero to $-1$.  Models with disk masses below
10\Mearth result in a constant dust production for $10^8$ years or
more, after which the curve turns into the common powerlaw with slope
$-1$. \newline
The speed at which collisional equilibrium is achieved is also
influenced by the size $a_c$ of the parent bodies
(Fig~\ref{fig:dependence}e).  If for a given disk mass, we reduce the
size of the parent bodies, the number of such particles will increase
and the collision time will become smaller.  Collisional equilibrium
is established after one or a few collision times. We can see this
effect clearly in the curves.  For the standard size of 1\km,
collisional equilibrium is reached after about $10^8$ years.  For
10\km{} bodies this takes $10^9$ years, and for 100\,m bodies,
collisional equilibrium is already fully established after $10^7$
years.\newline
Another very important parameter is the collision velocity
$\nu=\vcoll/\vkep$ (Fig.~\ref{fig:dependence}c).  Changing the
collision velocity changes the amount of dust seen in the system
significantly, in a counter-intuitive way: increasing the collision
velocity \emph{decreases} the amount of dust seen from a steady-state
cascade.  This behavior results from the fact that the \emph{grain
  removal} processes also strongly depend on the collisional
velocities.  Normally one would also expect that increasing the
velocity would lead to faster depletion of the cometary cloud.
However, this is not the case because the stirring of the orbits also
increases the volume in which the particles move (see
Eq.~\eqref{eq:2}) -- the collision times of comets are therefore not
influenced by the relative velocity.\newline
The distance of the cometary cloud (Fig~\ref{fig:dependence}c) to the
star also influences the observed \fdust values, in two ways.  First,
at a larger distance, the same amount of dust covers a smaller
fraction of the solid angle seen from the star.  The effect decreases
the \fdust value proportionally to $r^{-2}$.  Also, moving to larger
distances increases all time scales.  We can see in the figure that at
larger distances, it takes much longer for collisional equilibrium to
be established.  The time of almost constant \fdust is extended to
$10^9$ years if we move the dust production site from 43 to
150\AU.\newline
Finally, we can look at the dependence on the spectral type of the
star (Fig.~\ref{fig:dependence}f).  Moving from an A0 star to later
types slightly increases the amount of dust seen.  The most important
effect is that the size which dominates the visible dust will be
smaller for low-luminosity stars since the blow-out limit moves to
smaller grain sizes.  An additional small effect is the lower
mass of late-type stars which will reduce collision timescales.\newline
\cdrev{An additional parameter which influences the outcome of the
  calculations is the binding energy $S$, which is dependent on the
  internal composition and structure of the comets.  The binding
  energy determines the value of $\epsilon_0$ through Eqs.~\eqref{eq:37}
  and \eqref{eq:29}.  This parameter influences the speed at which
  material is processed through the collisional cascade and therefore
  the amount of visible dust.  The calculation shown in this paper all
  use $S=2\cdot10^6$\,erg/g, a value appropriate for icy comet-like
  bodies \citep[and references
  therein]{kenyon99:_accret_early_kuiper_belt}, and we assume this
  value to hold for the entire size range in the collisional cascade.
  Using $S=10^7$\,erg/g, a value more appropriate for asteroid-type
  bodies, table~\ref{tab:eps0} shows that the value of $\epsilon_0$
  would decrease by approximately a factor of 10.  Eqs.~\eqref{eq:24},
  ~\eqref{eq:260},\eqref{eq:30},\eqref{eq:32}, and \eqref{eq:34} all
  show a $\epsilon_0^{-1/2}$ dependence, so the corresponding
  timescales and masses as well as the amount of visible dust would
  increase by a factor of about 3.  Similarly, if the material the
  comets are made of would be exceptionally weak
  \citep{kenyon99:_accret_early_kuiper_belt}, the numerical values
  would decrease accordingly.}

An important result is: \emph{nowhere} in the parameter space covered
by Fig.~\ref{fig:dependence}, a slope of $t^{-2}$ is observed.  This
means that in all models shown here, the collisional removal of dust
grains dominates over the PR-drag.  This was to be expected from
equation~\eqref{eq:32} where we showed that the transition from
collisionally dominated dust removal to PR-drag dominated removal
processes happens only at rather low disk masses, which are
unobservable with current instrumentation.

\subsection{Summary of the collisional model}
\label{sec:summary}

\begin{enumerate}
\item Timescales do not depend critically on the collision velocities,
  if only the velocities are energetic enough to be destructive.
\item A powerlaw dependence of the amount of observed dust as a
  function of time can only be expected after about one collisional
  time for the bodies starting the cascade.  Before that time, an
  undisturbed cascade produces an approximately constant amount of
  dust.
\item For 1\,km-sized comets, the collision times are of order
  of $10^9$ years for disk masses (comets) of 1 Earth mass.  To reach
  powerlaw behavior within $10^8$ years and below, high disk masses
  ($10 \Mearth$) are required.  Alternatively, the collisional cascade
  could be started by 100 m bodies, provided that sufficient stirring
  can be achieved for these bodies.
\item An undisturbed collisional cascade predicts $\fdust=\mbox{const}$ for
  $t<\taucomet$ and $\fdust \propto t^{-1}$ at later times.
\item Disks in which PR drag dominates dust removal would show
  $\fdust\propto t^{-2}$ behavior.  However, the transition from
  collisionally dominated to PR drag dominated disks happens at disk
  masses of typically $10^{-3} \Mearth$, much less than required to
  support the observed debris disks.
\item Adding more mass to the disk does \emph{not} extend the lifetime
  of the debris disk.  For given collisional velocities, disks of
  different masses all converge (after one collision time \taucomet) to
  the same curve.  For more massive disks, this happens very quickly.
  The reason for this behavior is that in massive disks, the removal
  timescales are much shorter.  In fact, in a more complete model which
  also treats planet formation, massive disks would form planets
  quickly and remove comets by gravitational scattering, further
  shortening the lifetime of the debris disk.
\item The lower blowout sizes of late type stars help somewhat to
  increase the amount of dust seen around old stars, but not enough
  to explain the old Vega-like stars with significant debris disks.
\end{enumerate}

\section{Discussion}
\label{sec:discussion}

In this section we will relate the findings of the simple model to the
results of paper I, namely the main features of the distribution of
stars in the $\log t$ -- $\log \fdust$ diagram.

\subsection{The initial decrease for young stars}

\begin{figure}
\plotone{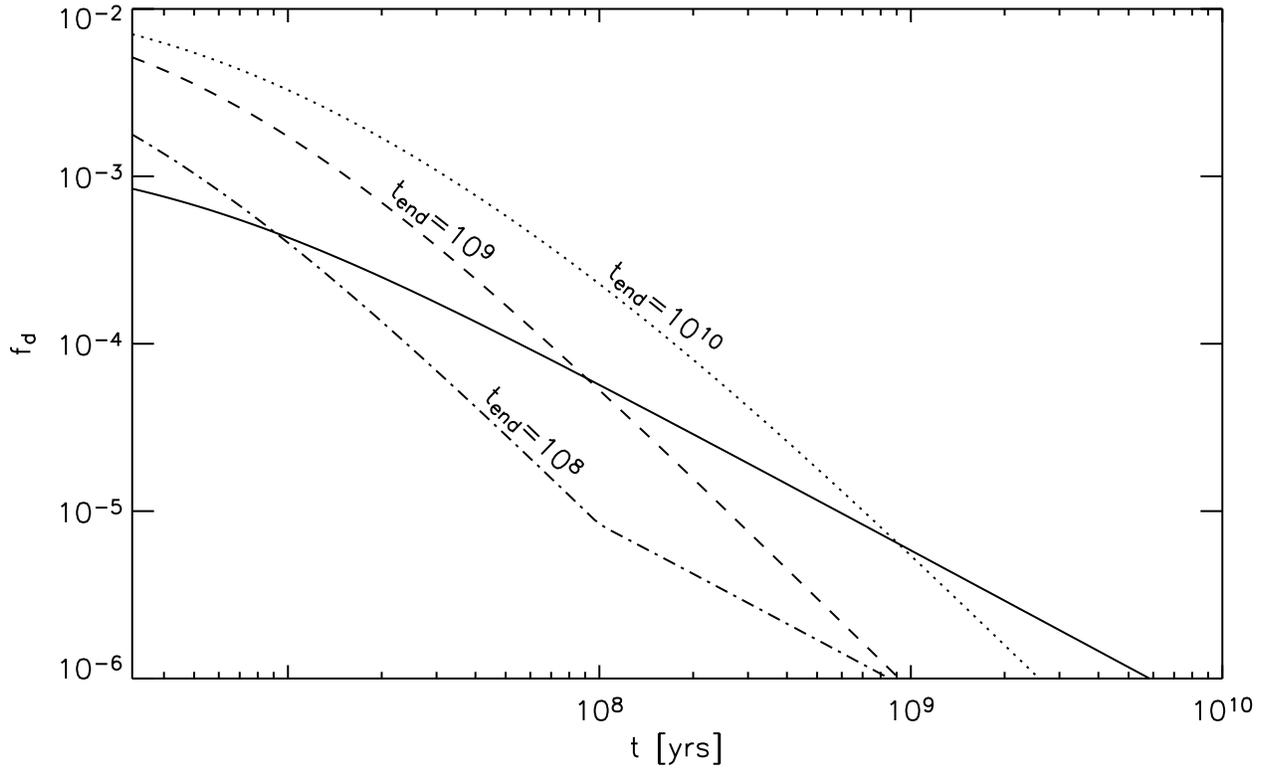}
\caption{\label{fig:stirr}\fdust-time relation with continuous stirring.
  Solid line:  the standard model without stirring.  The other lines
  include stirring, with $\nu$ starting at 0.01 and increasing
  linearly with time to a maximum value of 1.0 after 10$^8$, 10$^9$,
  and 10$^{10}$ years.}
\end{figure}

Looking at only the youngest stars in the sample of paper I, and
mainly at the A stars, there may indeed be an initial decrease in the
amount of dust present.  Due to the absence of very young stars with
low \fdust values, it seems clear that the stars at the edge of the
empty region in the lower left corner of the diagram must have evolved
from larger \fdust values at younger ages.  In this case, the lower
boundary of the data points would be marking out the fastest path of
decreasing the amount of dust in the system.
\citet{2001ApJ...555..932S} fitted a powerlaw with a slope $-1.7$ to
their sample, but did include T Tauri stars which are probably not
gas-free debris disks.  For the reduced sample, the slope of the lower
boundary is about $-1.3$, much closer to the slope of $-1$ which we
have derived for collisional cascades.  The data clearly is currently
not good enough to make a strong statement about the correctness of
this slope.  Hopefully, SIRTF will provide a much more solid database
for this study.  We would only like to make a remark here.  Suppose
the slope really is steeper than $-1$, what would that mean?  A pure
collisional cascade will not produce this.  However, we can see from
figure \ref{fig:dependence}, that an increase in collision velocity is
connected with a decrease in the \fdust value.  Therefore, a slope
steeper than $-1$ can be produced by a collisional cascade which is
\emph{continuously stirred}.  We have simulated this in a very simple
way, by increasing the collision velocity linearly in time starting
from an initial $\nu=0.01$ and ending at $\nu=1.0$ after a given
stirring time.  The result of this experiment can be seen in
Fig.~\ref{fig:stirr}.  Compared to the curve with a fixed $\nu=0.1$,
the stirring curves all show a steeper slope.

\placefigure{fig:stirr}

\subsection{The upper limit}
Vega-like stars never seem to have \fdust values larger than
$10^{-3}$.  This upper limit is independent of age, i.e. at all ages
do we find stars with $\fdust\sim 10^{-3}$, but not higher.  This
upper limit can be understood from the cascade dynamics.
Eq.~\eqref{eq:260} expresses the collision timescale for comets just
like Eq.~\eqref{eq:26}, but with the dependence on disk mass replaced
by \fdust.  From this equation we can see that the timescale for
survival of the cometary cloud drops below $10^8$ years when \fdust
exceeds $10^{-3}$.  An even stronger limit comes from work by
Artymowicz (\citeyear{1996rdfs.conf..137Asa}).  He showed that in a
gas-free debris disks, the amount of dust is limited by the creation
of \emph{dust avalanches}.  When \fdust reaches values of $10^{-2}$,
the break-up of particles in the inner regions of the disk creates
$\beta$-meteorites which are small particles driven out by radiation
pressure.  For low \fdust values, such particles will not suffer
additional collisions on the way out, because the disk is radially
optically thin. However, at \fdust values between $10^{-3}$ and
$10^{-2}$, the probability that such $\beta$-Meteorites will cause
additional collisions on their way out of the system becomes so high,
that a self-accelerating avalanche effect is created.  On basically a
Kepler time, the disk will be cleaned of small visible dust grains.
So even if the collisional equilibrium discussed in the present paper
does allow larger \fdust values for a limited period of time, the
avalanche mechanism effectively limits \fdust to about $10^{-3}$.
Stars with higher \fdust values must contain significant amounts of
gas.  The physics in such disks is different from ``normal'' Vega-like
systems, and studying the age dependence must therefore make a clear
distinction between the two types of disks.

\begin{figure}
\plotone{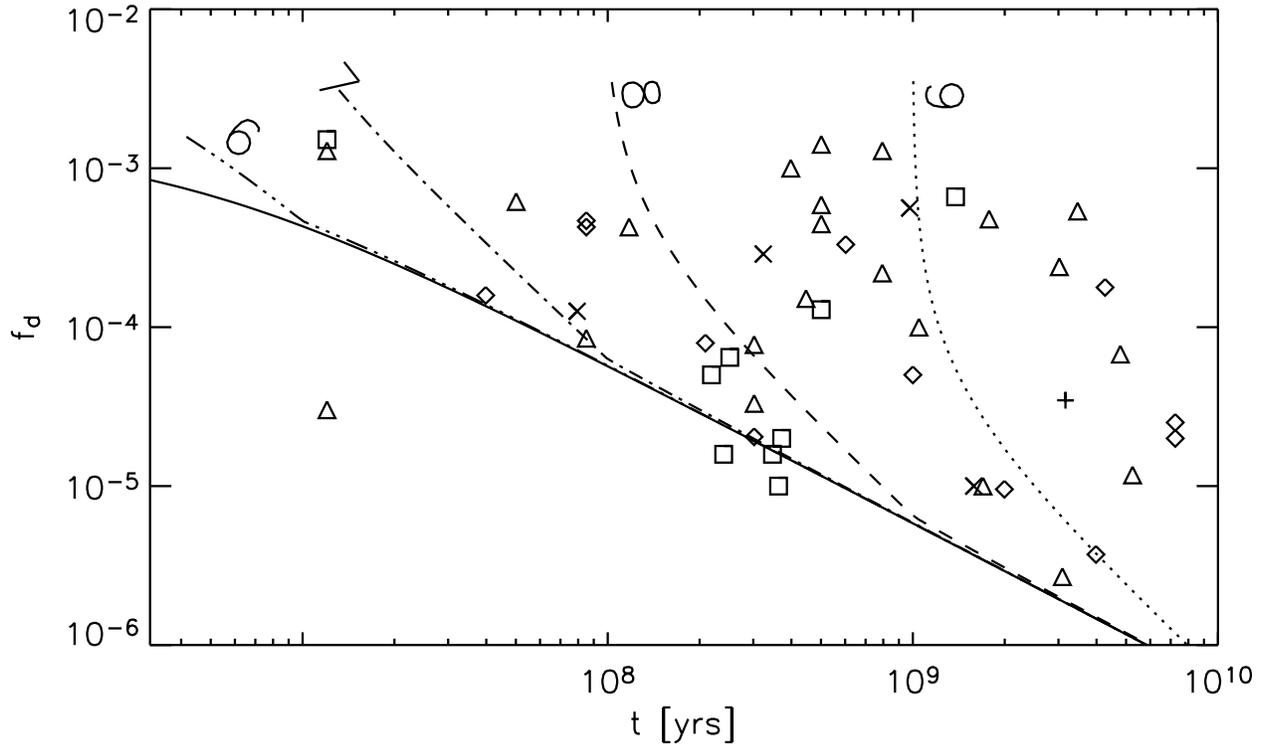}
\caption{\label{fig:stirr3}\fdust-time relation for different starting times of
  stirring.  Solid line: The standard model without stirring.  The
  other lines show models with different starting times for stirring.
  The starting time ($\log t_0$) is noted at the curve.  The stirring
  increases the velocity linearly between $t_0$ and $10\cdot t_0$
  where $t_0$ is $10^7$, $10^8$, and $10^9$ years.}
\end{figure}

\placefigure{fig:stirr3}

\subsection{Independence of the upper limit of the age}
Models of collisional cascades indicate, that the dust content of a
system should decrease with time.  This seems to be inconsistent with
the fact, that the observed debris disks can have $\fdust\simeq 10^{-3}$ at
all ages.  Assuming that the debris state starts at the same time for
all stars, a systematic decrease of dust abundance should result in an
absence of large \fdust values in old stars.  This is not what is
observed.  We have seen above, that a more or less constant amount of
dust can be supported before the disk goes into collisional
equilibrium, i.e. if the parent bodies are large, if the distance from
the star is large, or if the initial disk mass present in large
planetesimals is small.  However, figure~\ref{fig:dependence} also
shows that the types of solutions with constant dust for a long time
always are at levels significantly lower than $\fdust=10^{-3}$.
Therefore, a long duration of the debris state cannot be the
explanation for old Vega-like stars.  A more likely scenario is that
different planetesimal disks are starting to become active debris
disks at different times.  We can study this in a toy model, in which
we turn on the stirring of the planetesimal disk only after a given
waiting period.  Figure~\ref{fig:stirr3} shows the result of this
calculation.  In this way indeed solutions can be reached which cover
the observed points.  To find the reasons for the stirring, one needs
to run much more complete models which self-consistently include the
formation of planets.  Such models have recently been introduced by
\citet{2001AJ....121..538K,2002ApJ...577L..35K}.  They run a code
which computes the time evolution of the size distribution in an annulus
of a disk around a star.  Processes included are growth through
coagulation, collisional destruction, dynamical stirring and damping.
It is found that stirring by bodies of 500\,km or larger which form in
the disk can lead to the onset of a collisional cascade which will
produce debris dust \citep{2001AJ....121..538K}.  Furthermore, since
the timescales for the growth of larger bodies are larger in the outer
disk regions, planet formation and therefore debris production
proceeds from the inner to the outer disk in rings and can lead to multiple
collisional cascades at different times around the same star
\citep{2002ApJ...577L..35K}.  At least qualitatively, the ISO
observations and the estimates put forward in the present paper are
consistent with this picture.  The question remains, if this mechanism
can also explain very old Vega-like stars at ages of several Gyrs.

\subsection{The absence of young stars with low \fdust values}
The observations currently show no stars in this region.  We have
discussed in paper~I, that the samples are currently to small to
securely exclude the presence of stars in this region.  However, lets
for the moment take the relative absence of such stars as significant.
Then this indicates than normal debris disk do not populate this
region.  Looking at figure \ref{fig:dependence}, there are three main
ways to populate the area of young stars with low \fdust values.
\begin{enumerate}
\item\label{item:1} \textbf{Low initial disk masses below 10\Mearth.}\\
  Our collisional model produces low \fdust values for low initial
  disk masses.  However, the collisional velocity in the model is
  fixed.  Considering a self-stirring scenario as proposed by
  \citet{2002ApJ...577L..35K}, it is clear that enough mass must be
  present initially to allow for the formation of at least Pluto-like
  bodies.  \citet{2002ApJ...577L..35K} use a model with $100\Mearth$
  of solid material to reach the required velocities within 100 Myr.
  \citet{1997AJ....114..841S} find that at least 30\Mearth are needed
  in the early Kuiper Belt in order to grow Pluto.  \cdrev{Both
    calculations indicate that the time scale for planet formation and
    therefore for stirring is approximately inversely proportional to
    the disk mass.}  It is therefore likely that disks with less than
  about 10\Mearth of solid material in the Kuiper Belt region will not
  be able to start a collisional cascade in the first 100\,Myr because
  no large bodies will be formed in the belt.
\item  \textbf{High collision velocities very early on}\\
Low \fdust values can also be produced by very high collision
velocities in the disk, already very early, after about $10^7$ years.
However, such high velocities cannot be due to slow stirring by
Pluto-sized objects forming in the disk.  Large velocities are possible
by embedding a giant planet in the disk, but in this case all
planetesimals may have been removed by close encounters with the
planet within the first 10 Myrs.
\item \textbf{Collisional cascade at large distance from star}\\
  A collisional cascade at maybe 150\AU or further out seems to be
  capable of producing low \fdust values at young ages.  However, in
  this case the question must be raised: if there is enough mass in
  the disk at such large radii to produce stirring and a collisional
  cascade, why is there not also material closer to the star which
  would form planets faster, start the collisional cascade earlier
  and produce dust more efficiently?  If that additional material
  would be present, the star would become a ``normal'' debris disk
  with $\fdust\approx10^{-3}$.  The collisional cascade at large
  distances will then only dominate the IR output of the system after
  the initial debris ring closer to the star looses its capacity of
  producing dust because most planetesimals in this regions have been
  removed.
\end{enumerate}
Therefore, we speculate that the absence of debris disks with
intermediate or low \fdust values around \emph{young} stars is
\emph{due to the fact that stirring is only possible in  higher mass
  disks.}  Low mass disks, if they exist, will remain quiet for the
entire life of the star.

\section{Conclusions}

We have studied a simple model for collisional cascades in debris
disks.  While such a model does not lead to a self-consistent
description of the stirring and dust production in a debris disks, it
allows us to study the time behavior of such systems.  Comparing with
observations of debris disks, we come to the following conclusions.

\begin{enumerate}
\item\label{item:2} A collisional cascade with constant collision
  velocities leads to a powerlaw decrease of the amount of dust seen
  in a debris disk. The slope of that powerlaw is $-1$ for the
  parameter space valid for all observed debris disks.  Only at much
  lower masses (typically $10^{-3}\Mearth$) will this slope turn over
  to $-2$.
\item A collisional cascade which is continuously stirred (i.e. where
  the collision velocities are increasing with time) can produce
  slopes steeper than $-1$.
\item If the initial decrease of \fdust for young stars is confirmed
  by further observations, it may be due to a combination of stirring
  and collisional cascade, as described by
  \citet{2002ApJ...577L..35K}. 
\item The observed upper limit of $\fdust\simeq 10^{-3}$ for debris
  disks has to do with the dynamics of dust production in a collisional
  cascade and is due to the avalanche effect described by
  \citet{1996rdfs.conf..137Asa}.  According to that paper, larger
  \fdust values require gas to dominate the dust dynamics, but,
  following the definition by \citet{lagrange00:PPIV}, such disks are
  excluded from the class of debris disks.
\item The most likely explanation for the presence of debris disk with
  \fdust values up to $10^{-3}$ at ages well above a Gyr is the
  \emph{delayed onset of collisional cascades} by late planet formation
  further away from the star.  A prediction from this result is that
  the debris disks around older stars should be (on average) further
  away from the star than young debris disks.  
\item The tentatively observed absence of young debris disk with
  \fdust significantly lower than $10^{-3}$ may be real, and caused by
  the effect of stirring.  Low mass disks which could produce lower
  \fdust values cannot produce the planets needed to stir the disk
  quickly enough \citep{2002ApJ...577L..35K}.
\item The trends observed in debris disks so far need confirmation by
  a much larger sample, hopefully available after the launch of SIRTF.
\end{enumerate}

\acknowledgments

We are grateful to Rens Waters for inspiring discussions on the
subject.  GD is supported by project IUAP P5/36 financed by the
Belgian Federal Scientific Services (DWTC/SSTC).


\begin{thebibliography}{33}
\expandafter\ifx\csname natexlab\endcsname\relax\def\natexlab#1{#1}\fi

\bibitem[{{Artymowicz}(1988)}]{1988ApJ...335L..79A}
{Artymowicz}, P. 1988, \apjl, 335, L79

\bibitem[{{Artymowicz}(1996)}]{1996rdfs.conf..137Asa}
{Artymowicz}, P. 1996, in The Role of Dust in the Formation of Stars, ed. H.~U.
  {K{\" a}ufl} \& R.~Siebenmorgen (Berlin, Heidelberg, New York:
  Springer-Verlag), 137

\bibitem[{Aumann {et~al.}(1984)Aumann, Gillett, Beichmann, deJong, Houck, Low,
  Neugebauer, Walker, \& Wesselius}]{Aum-ea-84}
Aumann, H.~H., Gillett, F.~C., Beichmann, C.~A., deJong, T., Houck, J., Low,
  F.~J., Neugebauer, G., Walker, R., \& Wesselius, P.~R. 1984, \apjl, 278, 23

\bibitem[{Backman \& Paresce(1993)}]{backman-review}
Backman, D.~E., \& Paresce, F. 1993, in Protostars and Planets III, ed. E.~H.
  Levy \& J.~I. Lunine (Tuscon: University of Arizona Press), 1253--1304

\bibitem[{Burns {et~al.}(1979)Burns, Lamy, \& Soter}]{burns79:_radiat}
Burns, J.~A., Lamy, P.~L., \& Soter, S. 1979, Icarus, 40, 1

\bibitem[{{Davis} \& {Farinella}(1997)}]{1997Icar..125...50D}
{Davis}, D.~R., \& {Farinella}, P. 1997, Icarus, Vol. 125, p. 50-60 (1997),
  125, 50

\bibitem[{Decin {et~al.}(2003)Decin, Dominik, Waters, \& Waelkens}]{decin-I}
Decin, G., Dominik, C., Waters, L., \& Waelkens, C. 2003, \apj, in press

\bibitem[{{Dohnanyi}(1968)}]{1968IAUS...33..486D}
{Dohnanyi}, J.~S. 1968, in IAU Symp. 33: Physics and Dynamics of Meteors, 486

\bibitem[{{Durda} \& {Dermott}(1997)}]{1997Icar..130..140D}
{Durda}, D.~D., \& {Dermott}, S.~F. 1997, Icarus, 130, 140

\bibitem[{{Habing} {et~al.}(1999){Habing}, {Dominik}, {Jourdain de Muizon},
  {Kessler}, {Laureijs}, {Leech}, {Metcalfe}, {Salama}, {Siebenmorgen}, \&
  {Trams}}]{1999Natur.401..456H}
{Habing}, H.~J., {Dominik}, C., {Jourdain de Muizon}, M., {Kessler}, M.~F.,
  {Laureijs}, R.~J., {Leech}, K., {Metcalfe}, L., {Salama}, A., {Siebenmorgen},
  R., \& {Trams}, N. 1999, \nat, 401, 456

\bibitem[{{Habing} {et~al.}(2001){Habing}, {Dominik}, {Jourdain de Muizon},
  {Laureijs}, {Kessler}, {Leech}, {Metcalfe}, {Salama}, {Siebenmorgen},
  {Trams}, \& {Bouchet}}]{2001A&A...365..545H}
{Habing}, H.~J., {Dominik}, C., {Jourdain de Muizon}, M., {Laureijs}, R.~J.,
  {Kessler}, M.~F., {Leech}, K., {Metcalfe}, L., {Salama}, A., {Siebenmorgen},
  R., {Trams}, N., \& {Bouchet}, P. 2001, \aap, 365, 545

\bibitem[{{Holland} {et~al.}(1998){Holland}, {Greaves}, {Zuckerman}, {Webb},
  {MCCarthy}, {Coulson}, {Walther}, {Dent}, {Gear}, \&
  {Robson}}]{holland98:_submil}
{Holland}, W.~S., {Greaves}, J.~S., {Zuckerman}, B., {Webb}, R.~A., {MCCarthy},
  C., {Coulson}, I.~M., {Walther}, D.~M., {Dent}, W. R.~F., {Gear}, W.~K., \&
  {Robson}, I. 1998, \nat, 392, 788

\bibitem[{{Jourdain de Muizon} {et~al.}(1999){Jourdain de Muizon}, {Laureijs},
  {Dominik}, {Habing}, {Metcalfe}, {Siebenmorgen}, {Kessler}, {Bouchet},
  {Salama}, {Leech}, {Trams}, \& {Heske}}]{1999A&A...350..875J}
{Jourdain de Muizon}, M., {Laureijs}, R.~J., {Dominik}, C., {Habing}, H.~J.,
  {Metcalfe}, L., {Siebenmorgen}, R., {Kessler}, M.~F., {Bouchet}, P.,
  {Salama}, A., {Leech}, K., {Trams}, N., \& {Heske}, A. 1999, \aap, 350, 875

\bibitem[{{Jura} {et~al.}(1998){Jura}, {Malkan}, {White}, {Telesco}, {Pina}, \&
  {Fisher}}]{1998ApJ...505..897J}
{Jura}, M., {Malkan}, M., {White}, R., {Telesco}, C., {Pina}, R., \& {Fisher},
  R.~S. 1998, \apj, 505, 897

\bibitem[{{Kalas}(2000)}]{2000EM&P...81...27K}
{Kalas}, P. 2000, Earth Moon and Planets, 81, 27

\bibitem[{{Kenyon} \& {Bromley}(2001)}]{2001AJ....121..538K}
{Kenyon}, S.~J., \& {Bromley}, B.~C. 2001, \aj, 121, 538

\bibitem[{{Kenyon} \& {Bromley}(2002{\natexlab{a}})}]{2002AJ....123.1757K}
---. 2002{\natexlab{a}}, \aj, 123, 1757

\bibitem[{{Kenyon} \& {Bromley}(2002{\natexlab{b}})}]{2002ApJ...577L..35K}
---. 2002{\natexlab{b}}, \apjl, 577, L35

\bibitem[{{Kenyon} \&
  {Luu}(1999{\natexlab{a}})}]{kenyon99:_accret_early_kuiper_belt}
{Kenyon}, S.~J., \& {Luu}, J.~X. 1999{\natexlab{a}}, \aj, 118, 1101

\bibitem[{{Kenyon} \&
  {Luu}(1999{\natexlab{b}})}]{kenyon99:_accret_early_outer_solar_system}
---. 1999{\natexlab{b}}, \apj, 526, 465

\bibitem[{{Krivova} {et~al.}(2000){Krivova}, {Krivov}, \&
  {Mann}}]{2000ApJ...539..424K}
{Krivova}, N.~A., {Krivov}, A.~V., \& {Mann}, I. 2000, \apj, 539, 424

\bibitem[{{Lagrange} {et~al.}(2000){Lagrange}, {Backman}, \&
  {Artymowicz}}]{lagrange00:PPIV}
{Lagrange}, A.~., {Backman}, D.~E., \& {Artymowicz}, P. 2000, in Protostars and
  Planets IV, ed. V.~Mannings, A.~Boss, \& S.~S. Russell (Tucson: University of
  Arizona Press), 639

\bibitem[{{Laureijs} {et~al.}(2002){Laureijs}, {Jourdain de Muizon}, {Leech},
  {Siebenmorgen}, {Dominik}, {Habing}, {Trams}, \&
  {Kessler}}]{2002A&A...387..285L}
{Laureijs}, R.~J., {Jourdain de Muizon}, M., {Leech}, K., {Siebenmorgen}, R.,
  {Dominik}, C., {Habing}, H.~J., {Trams}, N., \& {Kessler}, M.~F. 2002, \aap,
  387, 285

\bibitem[{{Li} \& {Lunine}(2003)}]{2003ApJ...590..368L}
{Li}, A., \& {Lunine}, J.~I. 2003, \apj, 590, 368

\bibitem[{{Lissauer}(1993)}]{1993ARA&A..31..129L}
{Lissauer}, J.~J. 1993, \araa, 31, 129

\bibitem[{{Spangler} {et~al.}(2001){Spangler}, {Sargent}, {Silverstone},
  {Becklin}, \& {Zuckerman}}]{2001ApJ...555..932S}
{Spangler}, C., {Sargent}, A.~I., {Silverstone}, M.~D., {Becklin}, E.~E., \&
  {Zuckerman}, B. 2001, \apj, 555, 932

\bibitem[{{Stern} \& {Colwell}(1997{\natexlab{a}})}]{1997AJ....114..841S}
{Stern}, S.~A., \& {Colwell}, J.~E. 1997{\natexlab{a}}, \aj, 114, 841+

\bibitem[{{Stern} \& {Colwell}(1997{\natexlab{b}})}]{1997ApJ...490..879S}
---. 1997{\natexlab{b}}, \apj, 490, 879+

\bibitem[{{Tanaka} {et~al.}(1996){Tanaka}, {Inaba}, \&
  {Nakazawa}}]{1996Icar..123..450T}
{Tanaka}, H., {Inaba}, S., \& {Nakazawa}, K. 1996, Icarus, 123, 450

\bibitem[{{van der Hulst}(1981)}]{1981lssp.book.....V}
{van der Hulst}, H.~C. 1981, {Light scattering by small particles} (New York:
  Dover, 1981)

\bibitem[{{Williams} \& {Wetherill}(1994)}]{1994Icar..107..117W}
{Williams}, D.~R., \& {Wetherill}, G.~W. 1994, Icarus, 107, 117

\bibitem[{{Wyatt} {et~al.}(2003){Wyatt}, {Dent}, \&
  {Greaves}}]{2003MNRAS.342..876W}
{Wyatt}, M.~C., {Dent}, W.~R.~F., \& {Greaves}, J.~S. 2003, \mnras, 342, 876

\bibitem[{{Zuckerman} \& {Becklin}(1993)}]{zucker}
{Zuckerman}, B., \& {Becklin}, E.~E. 1993, \apj, 414, 793

\end{thebibliography}

\clearpage


\begin{table}
\caption{\label{tab:eps0} The $\epsilon_0$ parameter as a function of
  collision velocity $v_\mathrm{coll}$ and binding energy $S$.} 
\begin{center}
\begin{tabular}{r|rrrr}
\multicolumn{1}{c}{} & \multicolumn{4}{c}{$v_\mathrm{coll} \mbox{[km/s]}$} \\
$S \mbox{[erg/g]}$ & 0.01 & 0.1 & 1.0 & 10 \\\hline
$10^{4}$ &   $3.16\cdot 10^{1}$  &  $1.52\cdot 10^{3}$  &  $7.05\cdot 10^{4}$  &  $3.47\cdot 10^{6}$ \\
$10^{5}$ &   $2.80\cdot 10^{0}$  &  $2.22\cdot 10^{2}$  &  $1.04\cdot 10^{4}$  &  $4.81\cdot 10^{5}$ \\
$10^{6}$ &   $-$                 &  $3.16\cdot 10^{1}$  &  $1.52\cdot 10^{3}$  &  $7.05\cdot 10^{4}$ \\
$10^{7}$ &   $-$                 &  $2.80\cdot 10^{0}$  &  $2.22\cdot 10^{2}$  &  $1.04\cdot 10^{4}$
\end{tabular}
\end{center}
\end{table}

\clearpage

\clearpage 

\clearpage 

\end{document}